# On the way to quantum-gate implementation in permanently coupled AF spin rings without need of local fields

Filippo Troiani* and Marco Affronte

*INFM - National Research Center on nano-Structures and bio-Systems at Surfaces ($S^3$) and Dipartimento di Fisica, Università degli Studi di Modena e Reggio Emilia, Via Campi 213/A, 41100 Modena, Italy*

Stefano Carretta, Paolo Santini, and Giuseppe Amoretti

*INFM and Dipartimento di Fisica, Università di Parma, Parco Area delle Scienze, I-43100 Parma, Italy*

(Dated: February 24, 2005)

## Abstract

We propose a scheme for the implementation of quantum gates which is based on the qubit encoding in antiferromagnetic molecular rings. We show that a proper engineering of the intercluster link would result in an effective coupling that vanishes as far as the system is kept in the computational space, while it is turned on by a selective excitation of specific auxiliary states. These are also shown to allow the performing of single- and two-qubit gates without an individual addressing of the rings by means of local magnetic fields.

Localized electron spins in solid-state systems are widely investigated as potential building blocks of quantum devices and computers[1]. In fact, their relatively weak environmental coupling makes them ideal candidates for the realization of a long-lived quantum memory[2]. However, the spins of spatially separated electrons weakly interact also with each other: the implementation of the conditional dynamics thus requires either the use of auxiliary[3] or "bus" degrees of freedom[4], or additional resources (e.g., electric fields, more complex architectures) that allow a selective enhancement of the effective spin-spin interaction[5,6]. Alternative approaches are based on multi-spin encodings of the qubit in coupling-free subspaces, which render unnecessary the tuneability of the physical couplings during gating[7].

The implementation of single- (two-) qubit gates also requires the capability of selectively addressing single (couples of) spins by means a high spectral and/or spatial resolution. Though increased by the qubit encoding in spin clusters rather than single spins[8], the lenghtscales characterizing such local magnetic fields are still highly demanding from an experimental point of view. The global manipulation schemes[9–11], where quantum computation is performed also in the absence of such individual addressing, thus represent a valid alternative, maybe the most likely route in the nearest future.

In the present paper we propose a possible approach to the implementation of the single- and two-qubit gates, based on the qubit encoding in antiferromagnetic (AF) molecular rings[12] and on its manipulation by means of either local or global magnetic fields. In fact, the characteristic intracluster AF ordering of the spins is shown to allow an intercluster coupling which is effectively vanishing or finite, depending on the system being in its computational space or excited to specific auxiliary states, respectively. As an important point, this switching of the qubit-qubit coupling does not require additional resources, nor introduces longer timescales as compared to the ones that are used for performing the single-qubit gates. Moreover, the above scheme is shown to provide an efficient means for the implementation of a global-manipulation approach, already within a simple $ABAB\dots$ linear chain of rings and without requiring a multi-subsystem encoding of the qubit, nor the use of a large number of auxiliary units. *In both these respects, the existence of excited states that comes with the implementation of the qubit in spin clusters rather than single $s = 1/2$ spins represents a crucial resource.* The prototypical physical systems we have in mind are those substituted Cr-based AF rings characterized by the presence of a ground state doublet $S = 1/2$ which: (*i*) is energetically well separated from the excited levels, thus allowing to treat the molecule as an effective two-level systems; (*ii*) possesses a highly ordered, staggered arrangement of the spins, which would result in the effecive cancellation of suitably engineered intercluster couplings (see Ref. 13 and references therein). The scheme for the quantum-gate implementation, which can in principle be applied to analogous spin clusters as well, is first explained in general terms by referring to a model Hamiltonian, whose physical justification is given in the final part of the paper.

*Scheme for the quantum-gates implementation.* The two-qubit model Hamiltonian $\mathcal{H}^{AB} = \mathcal{H}^A + \mathcal{H}^B + \mathcal{H}^{AB}_{int}$ is defined in the Hilbert space of the effective two-level systems $A$ and $B$, namely $\{|0\rangle, |1\rangle, |i>1\rangle\}_A \otimes \{|0\rangle, |1\rangle, |j>1\rangle\}_B$, and reads as follows:

$$\mathcal{H}^{AB} = \sum_\alpha \sum_\beta (\epsilon^A_\alpha + \epsilon^B_\beta) |\alpha\rangle\langle\alpha| \otimes |\beta\rangle\langle\beta| + \sum_{\alpha,\beta>1} \mathcal{J}^{AB}_{\alpha\beta} |\alpha\rangle\langle\alpha| \otimes |\beta\rangle\langle\beta|, \qquad (1)$$

where the parameters $\mathcal{J}^{AB}_{\alpha\beta}$ are constant, reflecting the permanent nature of the intercluster coupling. As schematically shown in Fig. 1(a), $\mathcal{H}^{AB}_{int}$ leaves unaffected the energy levels corresponding to both $A$ and $B$ being in the respective computational spaces, while renormalizing the ones associated to the excited states of $B$. It should be noted that the above coupling is "asymmetric", for the excitation of $A$ to its own higher-lying states does not induce any renormalization in the energy levels as far as $B$ is in its computational space:

$$\mathcal{J}^{AB}_{\alpha\leq 1\,\beta>1} \neq 0, \quad \mathcal{J}^{AB}_{\alpha>1\,\beta\leq 1} = 0. \qquad (2)$$

These Hamiltonian and energy spectrum can be exploited for the implementation of the two-qubit gates in a straightforward manner. In fact, the application of a $\pi$-pulse centred on the frequency $\hbar\omega = \epsilon^B_{b_1} - \epsilon^B_1 + \mathcal{J}^{AB}_{1b_1}$ will only excite $B$ from $|1\rangle$ to the excited state $|b_1 > 1\rangle$ if $A$ is in its $|1\rangle$ state. A sequence of two such pulses, separated by a time delay $\tau$, would thus result in a conditional excitation and deexcitation, and more precisely in the application of a two-qubit quantum gate defined by the truth table $|\alpha\rangle \otimes |\beta\rangle \rightarrow e^{-i\alpha\beta\theta}|\alpha\rangle \otimes |\beta\rangle$, being $\alpha, \beta = 0, 1$ and $\theta = \tau(\epsilon^B_{b_1} + \mathcal{J}^{AB}_{1b_1} - \epsilon^B_1)/\hbar$. For $\theta = \pi$ the above transformation corresponds to a controlled-Z (CZ), that can be combined with two single-qubit rotations of the target qubit in order to perform a CNOT[14]. On the other hand, the unconditional dynamics (i.e., the single-qubit gates) requires the system to be kept in its computational space throughout the time evolution, for the transition energies between the $|0\rangle$ and $|1\rangle$ states of $A$ ($B$) doesn't depend on the setting of $B$ ($A$) (see Ref. 13 for a detailed discussion concerning the case of Cr$_7$Ni molecules). Therefore, the engineering of a time-independent two-qubit Hamiltonian $\mathcal{H}^{AB}$ such as the shown in Eq. 1, combined with a selective addressing of the spin clusters by means of local, pulsed magnetic fields, allows in principle to complete the set of universal quantum gates. The local manipulation of the qubits makes it superfluous (though not detrimental) for $A$ and $B$ to be physically different from each other.

The above effective Hamiltonian also allows an efficient implementation of quantum gates in the absence of a local control on the qubits. The kind of quantum hardware we will refer to in the following consists in a linear array $\ldots A_{n-1} B_{n-1} A_n B_n A_{n+1} B_{n+1} \ldots$ composed by two sets of identical molecules (Fig. 1(b)). These are assumed to differ from one another with respect to the energy eigenvalues, but to share the qualitative features of the low-energy spectrum. Correspondingly, the interaction between $A_n$ and its left neighbour $B_{n-1}$ is accounted for by an Hamiltonian $\mathcal{H}_{int}^{BA}$ which is identical to the $\mathcal{H}_{int}^{AB}$ of Eq. 1, except for the swapping of the molecules' labels. The implementation of quantum gates solely by means of global fields requires the use of auxiliary qubits in addition to the logical ones. In the following, the linear array will be assumed to consist in an alternated sequence of the two: the latter encode the quantum information, whereas the former are all set to $|0\rangle$, with the exception of the so-called "control unit" (CU), which is set to $|1\rangle$[9]. As demonstrated in Ref. 11, the capabilities required for the gates implementation with such a quantum hardware consist in: (i) the separate switching, on and off, of all the $A_n \leftrightarrow B_n$ and $B_{n-1} \leftrightarrow A_n$ couplings, and the application of $U_{CNOT}$ to the corresponding set of coupled cells; (ii) the performing of general rotations in each of the two subsets $A$ and $B$. More specifically, the single-qubit gates can be performed with an overall number of qubits which is twice that of the logical ones, whereas the complexity of the gating and the number of auxiliary qubits remarkably increase as far as the CNOT is concerned[15].

In the present case, the difference between the transition energies of $A$ and $B$ and the directionality of the effective coupling (Eq. 2) allow both the above conditions to be met. The conditional excitation of the molecules $A$ ($B$) to their auxiliary state $|a_1\rangle$ ($|b_1\rangle$) can be exploited for the implementation of the $U_{CZ}$ involving each $A_n$ ($B_n$) and its left neighbour $B_{n-1}$ ($A_n$), as shown in the previous paragraph (requirement (i)). For the interaction Hamiltonians $\mathcal{H}_{int}^{AB}$ and $\mathcal{H}_{int}^{BA}$ to be effectively turned off and the single-qubit rotations to be performed, instead, it suffices to manipulate the spin clusters within the respective computational spaces (requirement (ii)). Moreover, the existence of a futher excited state in at least one of the molecules (e.g., $|b_2 > 1\rangle$ in $B$) avoids the increase of additional auxiliary qubits for the CNOT implementation. In practice, the sequences of the SWAP gates that move the information within the array rigidly shift the states of the identical cells; being the logical qubits all encoded alternatively in $A$ or in $B$ and the physical interaction only effective between nearest neighbours, the main obstacle in performing the CNOT essentially consists in bringing the control- and the target-qubit next to each other. The specific resource that is provided by the present spin clusters is the above mentioned state $|b_2\rangle$: our aim is that of using $|b_2\rangle$ in order to block, and eventually invert, the motion of the qubits it collides with, thus provoking two or more logical qubits to be encoded on adjacent cells. In fact, the $U_{SWAP}$ rotation can be decomposed into a sequence of three $U_{CNOT}$[14]:

$$U_{SWAP}^{A_n B_n} = U_{CNOT}^{A_n B_n} \, U_{CNOT}^{B_n A_n} \, U_{CNOT}^{A_n B_n}, \qquad (3)$$

where $A_n$ ($B_n$) acts as a c-qubit (t-qubit) in the first and third $U_{CNOT}$, and as a t-qubit (c-qubit) in the second one. A $U_{CNOT}$ can in turn consist in the combination of a controlled-Z ($U_{CZ}$) and two rotations of the t-qubit, as shown in Fig. 2(a). As a key point, the initial setting of $B_n$ to its $|b_2\rangle$ state renders uneffective the transformations (i.e., puts out of resonance the transitions) enclosed in the dotted boxes; it is easy to verify that the remaining ones result in an identical transformation. In the same way, the application of $U_{SWAP}^{A_{n-1} B_n}$ does not change the state of $A_{n-1}$ (nor that of $B_n$) if $B_n$ is initially set to $|b_2\rangle$.

In Fig. 2(b) we show how to exploit this block for the implementation of a CNOT, where $|\phi_{n_0}\rangle$ and $|\phi_{n_0-1}\rangle$ act as a c- and a t-qubit, respectively. The required steps are the following: ($S_1$) after the CU has been positioned on the right-hand side of the c-qubit, a transformation $U_2^B \equiv |1\rangle\langle b_2| + |b_2\rangle\langle 1|$ is applied ($\pi$-pulse with central frequency $\hbar\omega = \epsilon_{b_2}^B - \epsilon_1^B$). As a result, the only $B$ cell that is excited to the auxiliary state $|b_2\rangle$ is the CU $B_{n_0}$, whereas all the others remain in the $|0\rangle$ state[16]; ($S_2$) a rotation $U_{SWAP}^{A_n B_n}$ swaps all the qubits but $A_{n_0}$ and $B_{n_0}$, thus allowing the collision of the control and target (logical) qubits, $|\phi_{n_0}\rangle$ and $|\phi_{n_0-1}\rangle$; a transformation ($S_3$) $U_{CNOT}^{A_n B_{n-1}}$ is performed, which is only effective on $|\phi_{n_0-1}\rangle$, being all the cells $A_{n \neq n_0}$ in their $|0\rangle$ state; ($S_4$) a second $U_{SWAP}^{A_n B_n}$, as in $S_2$, brings all the qubits back to their original positions; ($S_5$) a second $U_2 \equiv |1\rangle\langle b_2| + |b_2\rangle\langle 1|$, as in $S_1$, reinitializes the CU to its $|1\rangle$ state.

*Physical implementation of the qubit.* We now consider in greater detail the physical implementation of the qubit by referring to a class of prototypical systems, namely the substituted Cr$_x$Ni rings (with odd $x$), where the presence of the Ni ion typically results in the formation of a ground-state doublet. The precise identification of a spin cluster that meets all the above requirements is beyond the scope of the present paragraph, which is rather aimed at demonstrating the suitability of these specific molecules at a qualitative level, and at pointing out possible solutions for future molecular engineering. The single-qubit terms, which we have so far denoted by $\mathcal{H}^{X=A,B}$, correspond to the following spin Hamiltonian of the related $Cr_xNi$ molecule[12]:

$$\mathcal{H} = \sum_{i=1}^{x+1} J_i \mathbf{s}_i \cdot \mathbf{s}_{i+1} + \mu_B [B_0 \hat{\mathbf{z}} + \mathbf{B}_{xy}(t)] \cdot \sum_{i=1}^{x+1} g_i \mathbf{s}_i + \sum_{i=1}^{x+1} d_i [(s_{z,i})^2 - s_i(s_i+1)/3] + \sum_{i<j=1}^{x+1} \mathbf{s}_i \cdot \mathbf{D}_{ij} \cdot \mathbf{s}_j, \quad (4)$$

$J_{i=2-x+1} = J$ and $J_{x,x+1} = J'$, $g_{i=1-x} = g$ and $g_{x+1} = g'$, $s_{i=1-x} = 3/2$ and $s_{x+1} = 1$, being $i = x + 1$ the site of the Ni substitution. As discussed in detail in Ref. 13 for the case $x = 7$, the ground-state doublet that forms the computational basis $\{|0\rangle, |1\rangle\}$ approximately corresponds to $\{|S = 1/2, S_z = -1/2\rangle, |S = 1/2, S_z = +1/2\rangle\}$; these states are characterized by an antiparallel alignment of the neighbouring spins, which we propose to exploit in order to obtain the effective cancellation of the intercluster coupling. The auxiliary states $|b_1\rangle$ and $|b_2\rangle$ are respectively identified with the first excited states $M = -3/2$ and $M = -1/2$ ($b_{1,2} = 2, 3$), belonging to the multiplet $S = 3/2$.

The realization of an intercluster coupling has been demonstrated in similar systems[17] (ferromagnetic spin clusters), as well as quantum-mechanical entanglement[18]. For the sake of clarity, we shall assume in the following that an effective coupling between the rings $A$ and $B$ may result from one or more links that connect individual spins belonging to the two of them. An Ising-like interaction between spin $i$ of cluster $A$ and spin $j$ of cluster $B$, namely $\mathcal{H}_{int}^{AB} = J_{ij}^{AB} s_{z,i}^A s_{z,j}^B$, results in an analogous coupling between $A$ and $B$:

$$\mathcal{J}_{\alpha\beta}^{AB} = \frac{J_{ij}^{AB} \langle \alpha || s_{z,i}^A || \alpha \rangle \langle \beta || s_{z,j}^B || \beta \rangle S_{z,\alpha}^A S_{z,\beta}^B}{\sqrt{S_A(S_A+1)(2S_A+1)S_B(S_B+1)(2S_B+1)}}, \quad (5)$$

where $S_{z,\alpha}^A (S_{z,\beta}^B)$ is the average value of $S_z^A (S_z^B)$ corresponding to state $|\alpha\rangle$ of $A$ ($|\beta\rangle$ of $B$), and $\langle 0 || s_{z,i}^{A,B} || 0 \rangle = \langle 1 || s_{z,i}^{A,B} || 1 \rangle$, being the reduced matrix elements constant within each $S$-multiplet. In presence of more than one spin-spin interaction, Eq. (5) can be generalized by summing over the $i$ and $j$ indexes. In particular, we will consider the case where the spin $i$ of cluster $A$ is linked to two adjacent spins, $j$ and $j+1$, of $B$, with $J_{ij}^{AB} = J_{i\,j+1}^{AB}$ (see Fig. 1(c)).

TABLE I: Reduced matrix elements $\langle\beta||s_{z,k}||\beta\rangle$ in the Cr$_5$Ni (up), Cr$_7$Ni (middle) and Cr$_9$Ni (down) rings. The index values $k=6$ (up), k=8 (middle), and $k=10$ (down) correspond to the Ni substitution.

| $\beta$ | $s_{z,1}$ | $s_{z,2}$ | $s_{z,3}$ | $s_{z,4}$ | $s_{z,5}$ | $s_{z,6}$ | $s_{z,7}$ | $s_{z,8}$ | $s_{z,9}$ | $s_{z,10}$ |
|---|---|---|---|---|---|---|---|---|---|---|
| 0,1 | 1.48 | -1.17 | 1.40 | -1.17 | 1.48 | -0.80 | - | - | - | - |
| $b_1,b_2$ | 1.51 | -0.13 | 1.28 | -0.13 | 1.51 | -0.17 | - | - | - | - |
| 0,1 | 1.39 | -1.13 | 1.30 | -1.12 | 1.30 | -1.13 | 1.39 | -0.77 | - | - |
| $b_1,b_2$ | 1.40 | -0.30 | 1.09 | -0.11 | 1.09 | -0.30 | 1.40 | -0.39 | - | - |
| 0,1 | 1.33 | -1.10 | 1.23 | -1.08 | 1.21 | -1.08 | 1.23 | -1.10 | 1.33 | -0.75 |
| $b_1,b_2$ | 1.31 | -0.44 | 1.01 | -0.21 | 0.90 | -0.21 | 1.01 | -0.44 | 1.31 | -0.38 |

As a consequence of the approximately staggered (AF) arrangement of the reduced matrix elements characterizing $|\beta=0,1\rangle$ (see Table I and relative discussion), the sum of the two individual couplings results in a cancellation of the effective one between the rings:

$$\langle\beta||s_j^B||\beta\rangle = -\langle\beta||s_{j+1}^B||\beta\rangle \implies \mathcal{J}_{\alpha\beta}^{AB} = 0 \text{ for } \beta=0,1. \tag{6}$$

On the other hand, the excitation of $B$ to a state lacking such intracluster AF arrangment switches on the effective coupling to $A$ (see Fig. 1(d)). It should be noted that the key result, namely Eq. 6, doesn't depend neither on the assumed Ising-like form of $\mathcal{H}_{int}^{AB}$, nor on the presence of two disinct links: an Heisenberg coupling or a single link symmetrically coupled to $s_{z,j}^B$ and $s_{z,j+1}^B$ would lead to the same result.

In order to verify to which extent Eq. 6 holds in realistic Cr$_x$Ni rings, we have performed detailed calculations for $x=5,7,9$ (Table I). While a relevant increase is actually achieved by exciting the system from $|0\rangle$ or $|1\rangle$ to $|b_1\rangle$, the vanishing of $\mathcal{J}_{\alpha\beta}^{AB}$ in the computational basis would require slightly asymmetric links, such that $J_{i,j}^{AB}/J_{i\,j+1}^{AB} = -\langle\beta||s_{j+1}||\beta\rangle/\langle\beta||s_j||\beta\rangle = |\langle\beta||s_{j+1}||\beta\rangle/\langle\beta||s_j||\beta\rangle|$ ($\beta=0,1$).

As already mentioned, a crucial role within our scheme is played by the (excited) auxiliary states of the molecules. The same possibility of inducing direct transitions between these and the ground-state doublet by means of an oscillating magnetic field is a non-trivial issue, for it requires non-vanishing matrix elements $\langle\beta|\sum_{i=1}^{x+1} g_i \mathbf{s}_i|\beta'\rangle$, with $\beta=0,1$ and $\beta'=b_1,b_2$. In fact, in Cr$_7$Ni these are approximately proportional to $g_{Ni}-g_{Cr} \simeq 2.2-1.98$, and are thus finite only due to the presence of the Ni substitution; besides, they depend on the value of the static magnetic field $B_0$ (for $B_0 = 2$ T[13], for example, $|\langle 0| \sum_{i=1}^{8} g_i s_{x,i} |1\rangle / \langle 0| \sum_{i=1}^{8} g_i s_{x,i} |b_1\rangle| = 3.8$). From these calculations, and from others performed on analogous systems (e.g., Cr$_7$Fe and Ni$_7$Cu), we conclude that, as far as the impurity-Cr exchange is comparable to the Cr-Cr one, the main physical properties required by the discussed approach are satisfied. Therefore, a possible implementation of the linear $A$-$B$ array[19] may be composed by Cr$_x$Ni rings with two different values of Cr ions ($x_A \neq x_B$), or by two among the existing variants a given of Cr$_x$Ni ($x_A = x_B$), characterized by different values of the physical parameters $J$ and $g$.

In conclusion, we have proposed a scheme for the implementation of quantum gates in AF spin clusters which does not require the intercluster interaction to be tuneable during gating. Having in mind the prototypical case of the Cr-based rings, we have shown how specific auxiliary states allow in principle to control the effective coupling between the qubits, and to implement global-manipulation approaches, with a minimum number of auxiliary qubits, smaller than the one required by single-spin encodings.

---

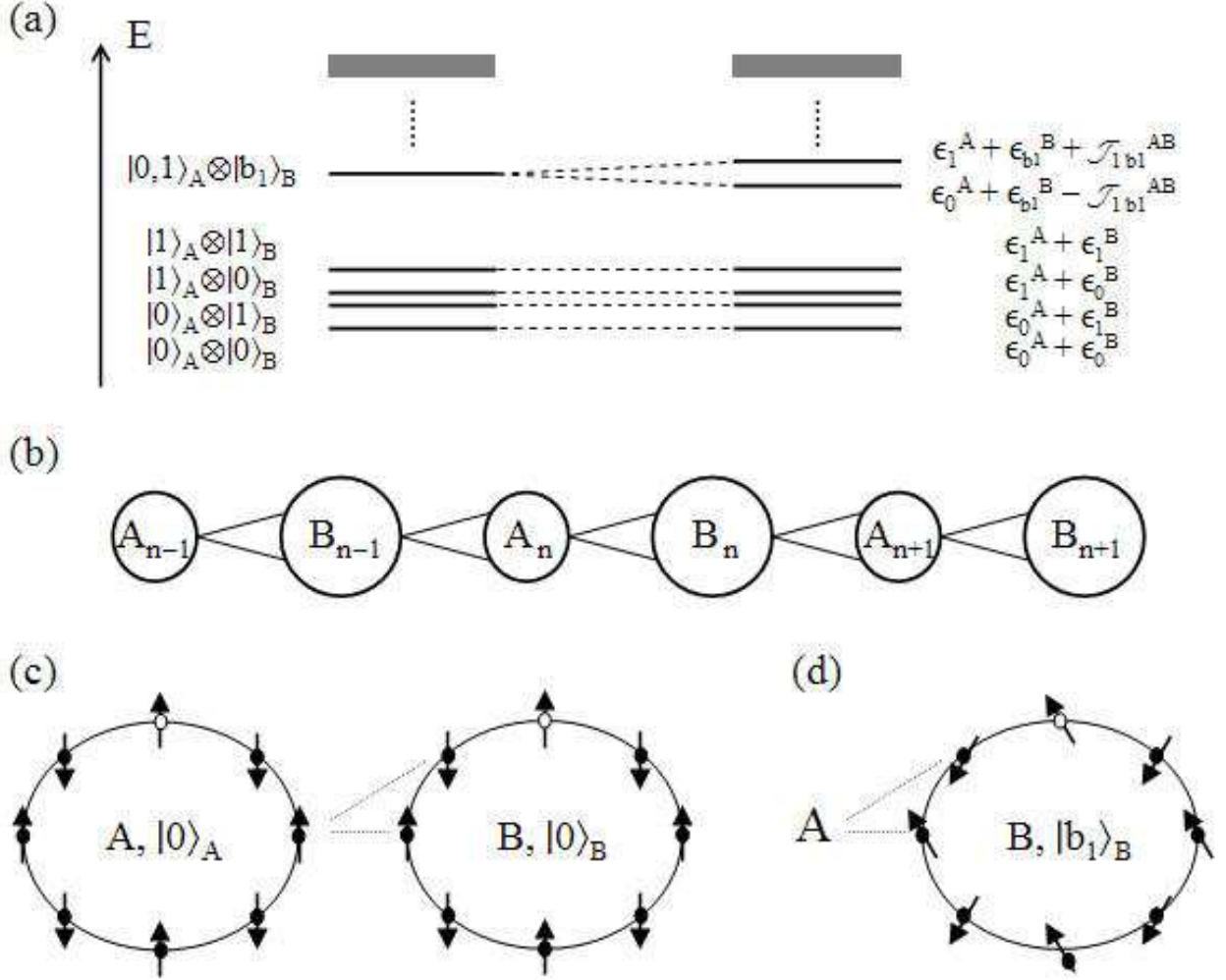

FIG. 1: (a) Schematics of the lowest energy levels of the $AB$ system with and without (left- and right-hand sides, respectively) the effective intercluster coupling. The levels corresponding to the computational space do not undergo energy renormalizations; being $S^A_{z,0} = -S^A_{z,1}$, $\mathcal{J}^{AB}_{0\,b_1} = -\mathcal{J}^{AB}_{1\,b_1}$ (see Eq. 5). (b) Sketch of the linear array of rings $ABAB\ldots$, with asymmetric couplings between each two neighbouring units. (c) Intracluster ordering of the electron spins characterizing the ground-state doublet of a prototypical AF ring, namely the substituted Cr$_7$Ni. Possible intercluster molecular links (solid lines), locally coupling spin $i$ of ring $A$ with $j$ and $j+1$ of $B$. (d) The excited state $|b_1\rangle$ of $B$ lacks the above antiparallel arrangement of the adjacent spins.

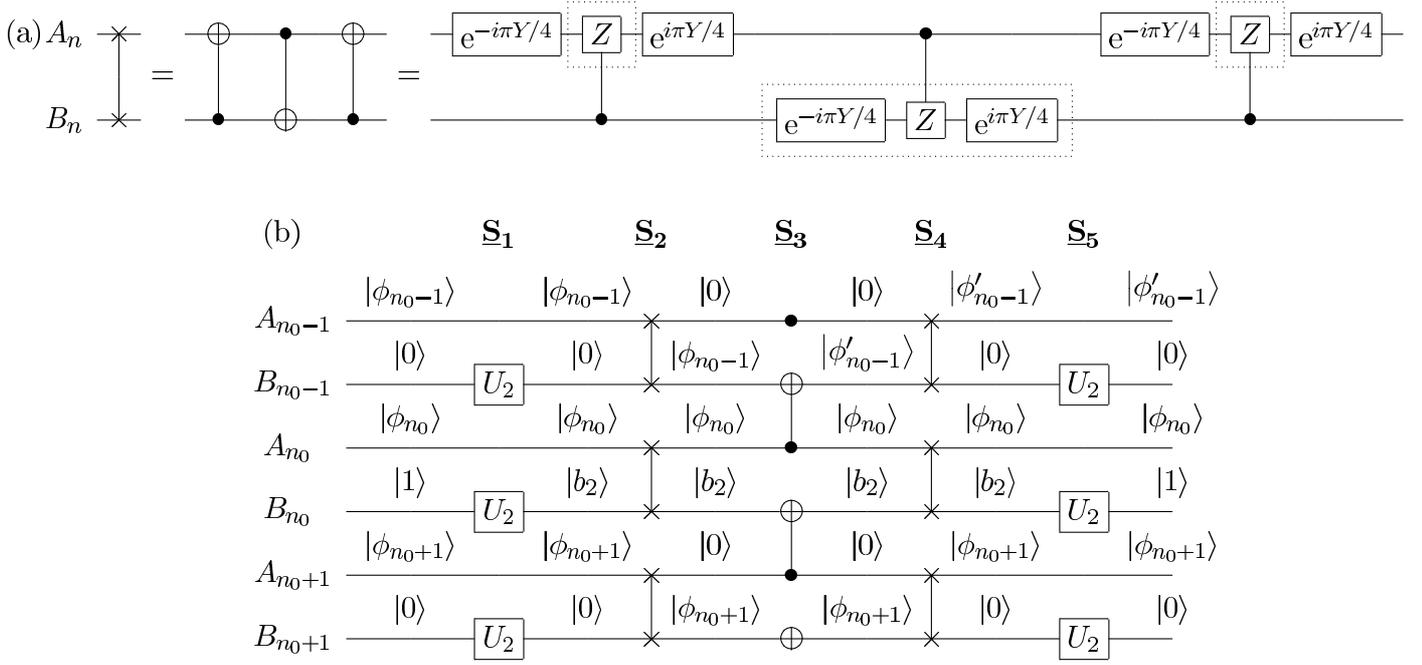

FIG. 2: (a) Decomposition of $U_{SWAP}^{A_nB_n}$ in terms of $U_{CZ}$ and single-cell rotations ($Y$ and $Z$ correspond to the Pauli matrixes $\sigma_y$ and $\sigma_z$, respectively). The rotations enclosed in the dotted boxes are uneffective if $B_n$ is initially set outside from its computational space, to the auxiliary state $|b_2\rangle$. The resulting unitary transformation can thus be written as $U_{SWAP}^{A_nB_n} = |00\rangle\langle00|+|11\rangle\langle11|+|01\rangle\langle10|+|10\rangle\langle01|+(|0\rangle\langle0|+|1\rangle\langle1|)\otimes|b_2\rangle\langle b_2|$. (b) CNOT gate applied in five steps ($S_{1-5}$) to $|\phi_{n_0}\rangle$ and $|\phi_{n_0-1}\rangle$, acting as control- and target-qubit, respectively. On top of the wires we report the quantum states of the cells after each step (for simplicity, such notation refers to the case where the subsystems are never entangled with each other; we can thus write $|\phi_{n_0}\rangle \otimes |\phi'_{n_0-1}\rangle = U_{CNOT}|\phi_{n_0}\rangle \otimes |\phi'_{n_0-1}\rangle$).